\documentclass[12pt]{article}
\usepackage{epsfig}

\def\beq{\begin{equation}}
\def\eeq#1{\label{#1}\end{equation}}
\def\eeqn{\end{equation}}

\def\beqa{\begin{eqnarray}}
\def\eeqa#1{\label{#1}\end{eqnarray}}
\def\eeqan{\end{eqnarray}}

\let\bar=\overbar

\def\Dslash{\not{\hbox{\kern-4pt $D$}}}
\def\dslash{\not{\hbox{\kern-2pt $\del$}}}

\def\msb{{\bar{\ssstyle M \kern -1pt S}}}

\usepackage{fancyhdr,graphicx}
\def\Title#1{\begin{center} {\Large {\bf #1} } \end{center}}

\begin{document}

\Title{Statistical measure of complexity in compact stars with global charge neutrality}

\bigskip\bigskip

\begin{raggedright}
{\it Rodrigo A de Souza $^{1}$\\  Marcio G B de Avellar $^{2}$ \\ Jorge E Horvath $^{3}$\index{de Souza, R.}\\
Instituto de Astronomia, Geof\'isica e Ci\^encias Atmosf\'ericas\\
Universidade de S\~ao Paulo\\
05570-010 Cidade Universit\'aria\\
S\~ao Paulo, SP\\
Brazil\\
{\tt Email: rodrigo.souza@usp.br $^{1}$ \\ \hspace{1.75cm}marcavel@astro.iag.usp.br $^{2}$ \\ \hspace{1.75cm}foton@astro.iag.usp.br $^{3}$}}
\bigskip\bigskip
\end{raggedright}

\section{Introduction}

In the past three decades, information theoretic methods have been applied to many different systems, from molecular biology to 
quantum mechanical systems, and even to linguistics. Specifically speaking about quantum systems, these methods can reveal the presence 
of interactions, correlations of experimentally measured quantities, universal relations and much more.

The basic concept of information theory is the Shannon Information, also known as Shannon Entropy or Information Entropy. But the 
question arises: what exactly is information? Information, in a general sense, is whatever we get about the occurrence of a given event: 
for exemple, how surprising or unexpected results. Shannon defined \cite{shannon48}, in 1948, an expression that measures the information (or randomness, 
or uncertainty, or ignorance) about a system. Obeying a set of mathematical properties defined by him (and even if subject to a certain 
reductionism), the information content of a system in terms of probabilities of a event to occur is:

\begin{equation}
H=-K\sum_{i}p_{i}log_{b}[p_{i}]\hspace{0.5cm}or\hspace{0.5cm}H=-K\int p(x)log_{b}[p(x)]dx,
\end{equation} respectively for the discrete and continuous cases.

From this, people started thinking about how complex a system can be and how to measure this complexity calculating it from a mathematical 
definition. Thus, the statistical measure of complexity introduced by Lopez-Ruiz, Mancini and Calbet \cite{lopezruiz95} relates the complexity of a system 
to the information stored in it and the distance to a situation in which all possible states of the system are equiprobable. This definition 
encodes the concepts of order and disorder of a given arrangement of the system.

To illustrate these concepts we may think about two ideal systems frequently used in physics: the ideal gas and the perfect crystal, extremes 
in all aspects and opposites as well. Because they both are idealized systems, they should be thought as minimally complex systems. However, 
while the latter is totally ordered, the former is totally disordered; i.e. while for the perfect crystal one state is more probable than the 
others, in the ideal gas all states are equiprobable. Summarizing this intuitive view:

\begin{itemize}
	\item Perfect crystal: This system has zero complexity by definition; its strict symmetry rules implies probability density centered 
	around the prevailing state of perfect symmetry which result in minimal information. The system is completely ordered.
	\item Ideal gas: Ideal gases have also zero complexity by definition; Their accessible states are equiprobable resulting in maximal 
	information. The system is totally disordered.
\end{itemize}

\begin{figure}[htb]
\begin{center}
\epsfig{file=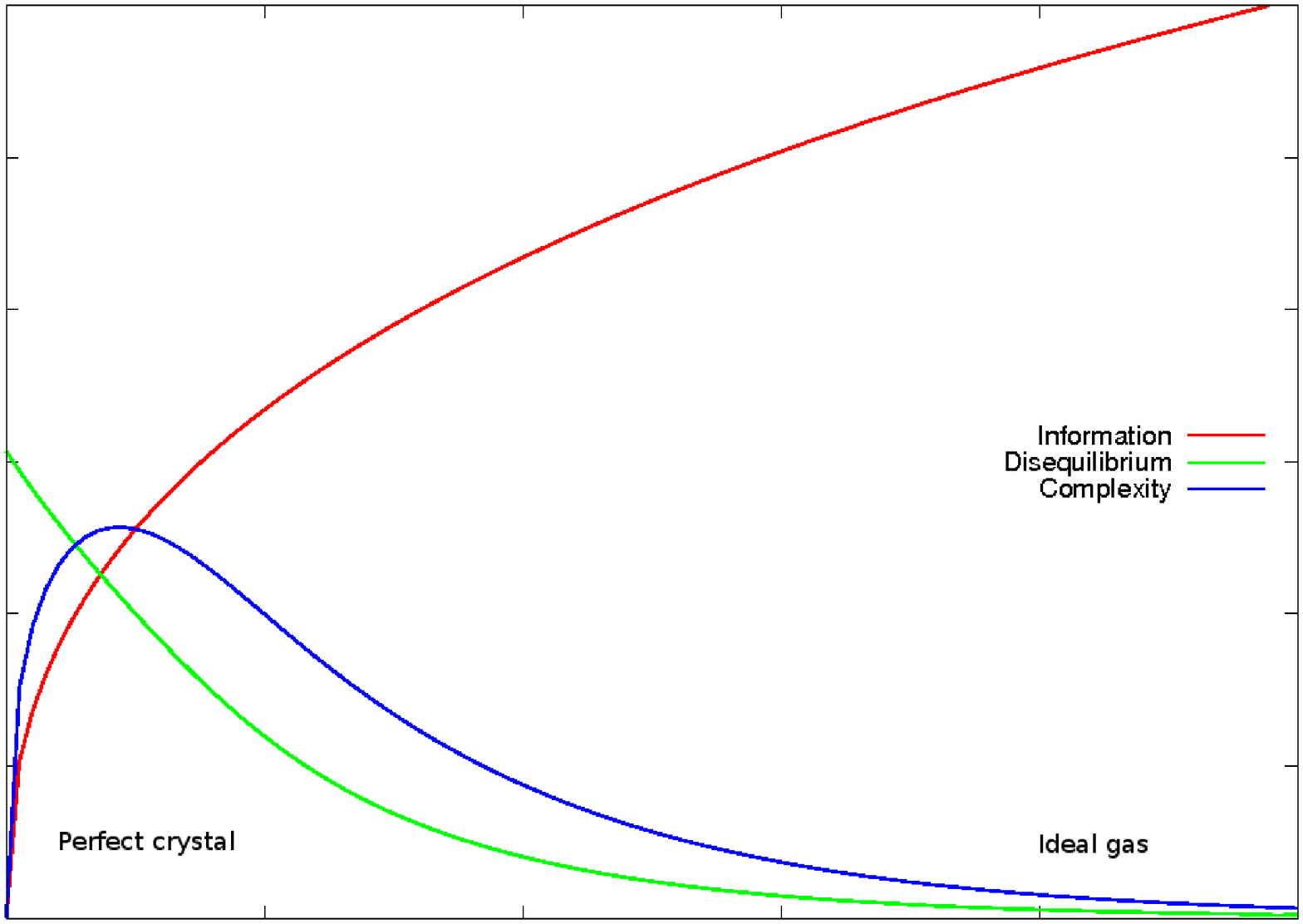,height=3in}
\caption{Intuition of what should be complexity, at least asymptotically.}
\label{fig:intuition}
\end{center}
\end{figure}

The intuitive behavior/relation among complexity, information entropy and disequilibrium is expected to be the one shown in Figure \ref{fig:intuition}.

At this stage an suitable expression for complexity arises:

\begin{equation}
C\equiv H\times D\hspace{0.5cm}or\hspace{0.5cm}C\equiv e^{H}\times D,
\end{equation} where

$$
D=\sum_{i}\Big[p_{i}-\frac{1}{N}\Big]^{2}\hspace{0.5cm}or\hspace{0.5cm}D=\int p^{2}(x)dx
$$ is the distance to the (equilibrium) equiprobability of states.

In this work, we used the measure of complexity given by $C=e^{H}D$ to study the effect of the global charge neutrality developed by Rotondo 
et al \cite{rotondo11} on neutron star structure following the work by de Avellar and Horvath \cite{avellar12} who compared the complexity of sequences of stars 
with different equations of state using the ``standard'' local charge neutrality.

The standard approach to the neutron star structure assumes the relativistic hydrostatic equilibrium condition and a (realistic) equation 
of state. Hidden in this scheme is, by construction, the assumption of local charge neutrality leading to no global electric field inside the star. However, recent theoretical developments \cite{ruffini10} concluded that the insurgence of a critical electrical field during the gravitational collapse leads to the necessity of a full reexamination of the gravito-electrodynamical properties of neutron stars. If this is true, then one needs to consider an extension of the $\beta$-equilibrium condition consistently within, for example, a relativistic Thomas-Fermi equation, otherwise there could not be an equilibrium on microphysical scales.

Thus, when constructing the mathematical equations for the structure of these compact objects one needs to couple the relativistic 
Thomas-Fermi equation with the equilibrium condition governed by the Einstein-Maxwell equations.

In an extension of the work quoted in \cite{rotondo11}, Belvedere et al \cite{belvedere12} included the strong interaction making the coupled set of equations even 
more difficult to solve analytically.

Here we intend to study first the effects of the global charge neutrality in the order of the system and then to study the effects of 
the inclusion of the strong interaction. Our study potentially leads to the construction of a hierarchy of equations of state to be realized 
in nature from the informational theoretic methods.

\section{Results and Conclusions}

Our results show the preliminary calculations for the two density profiles from reference \cite{belvedere12} (Table \ref{tab:profileCalculations}). It seems that the global charge neutrality and 
the presence of strong interactions actually lower the disequilibrium of the star sequence in a way that the star tend to the ideal gas case 
in our intuition plot.

\begin{table}[h!]
\begin{center}
\begin{tabular}{|c|c|c|c|c|c|}
\hline
Neutrality & $M [M_{\odot}]$ & $R [km]$ & $H [nats]$ & $D$ & $C$ \\
\hline
\multicolumn{6}{|c|}{$\rho_{crust}=10^{10} g/cm^{3}$} \\
\hline
Global & 2.0356 & 12.3386 & -0.586 & 30.441 & 16.947 \\
Local & 2.2354 & 13.4787 & -0.840 & 36.611 & 15.583 \\
\hline
\multicolumn{6}{|c|}{$\rho_{crust}=4.3\times10^{11} g/cm^{3}$} \\
\hline
Global & 1.8707 & 12.5156 & -0.488 & 23.583 & 14.474 \\
Local & 1.9794 & 13.3375 & -0.768 & 29.866 & 13.858 \\
\hline
\end{tabular}
\caption{Statistical measures of the information content, disequilibrium and complexity for two density profiles from ref. \cite{belvedere12}.}
\label{tab:profileCalculations}
\end{center}
\end{table}

If our full calculations validate this results to the entire sequence of stars, we could have a direct measure of the effects of the global 
charge neutrality via information theoretic methods on the structure of neutron stars making these interactions and conditions more probable 
to be realized in nature.

\section{Perspectives and further developments}

Besides the total implementation of the code to solve the structure of neutron stars with local and global charge neutrality in order to study 
the information content of the different equations of state, we are developing the theory to further validate the use of the density profile as 
a probability-like distribution to be used in the calculation of the information content of a system, avoiding the ``negative'' information as 
done so far. In particular, we defined a new density profile satisfying two important features of probability functions:

\begin{enumerate}
\item p(x) $\in$ [0:1] $\forall$ $x$,
\item $\int p(x)dx = 1$.
\end{enumerate}

Using a exact density profile from a well-known exact solution of the Einstein equations we could match these conditions, yielding the results shown in 
Figure \cite{fig:information_content}.

\begin{figure}[htb!]
\begin{center}
\epsfig{file=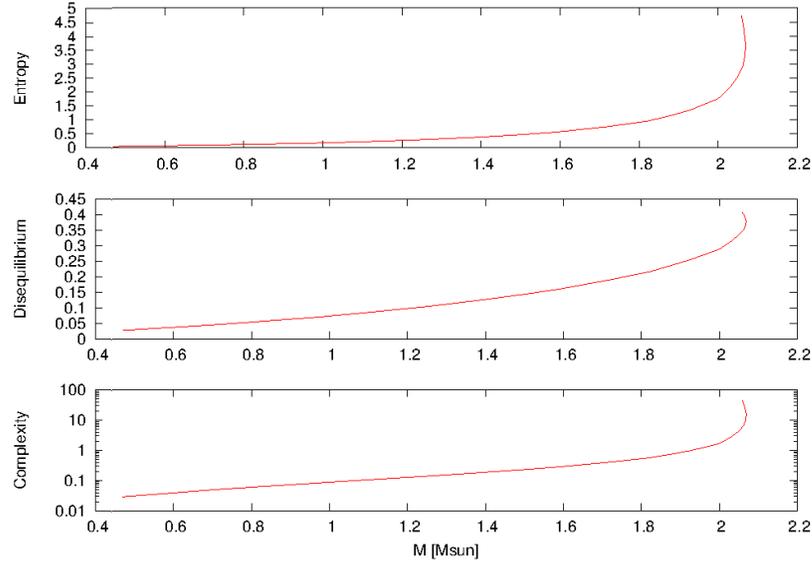,height=3in}
\caption{Information content, disequilibrium and complexity of the sequence of stars from the exact solution of the TOV equations 
taking into account anisotropy in pressure and MIT Bag Model equation of state [and local charge neutrality]}
\label{fig:information_content}
\end{center}
\end{figure}

However, the consistency conditions alone are not enough. It is very necessary to understand better the meaning of what we want to calculate. 
The results for complexity shown in Figure \ref{fig:information_content} are at odds with the conclusion by de Avellar and Horvath \cite{avellar12} and by Chatzisavvas et al. \cite{chatzisavvas09}, who 
stated that ``neutron stars are ordered systems that cannot grow in complexity as its mass increases''. The reason behind this difference lies in the 
rate at which $e^{H}$ increases or decreases relatively to the disequilibrium $D$ and this, in turn, is related to the signal of $H$. Thus, 
further studies 
along these lines are required to characterize how these quantities behave for self-gravitating stars.


\begin{thebibliography}{99}

\bibitem{shannon48} Shannon, C. E., {\it Bell System Technical Journal}, 27, 379-424, 623-656, 1948.

\bibitem{lopezruiz95} L\'opez-Ruiz, R.; Mancini, H. L. and Calbet, X., {\it Physics Letters A}, 1995, 209, 321-326

\bibitem{rotondo11} Rotondo, M., Rueda, J. A., Ruffini, R., Xue, S.-S. {\it Physics Letters B}, 2011, 701, 667-771

\bibitem{avellar12} de Avellar, M. G. B., Horvath, J. E., {\it Physics Letters A}, 2012, 376, 1085-1089

\bibitem{ruffini10} Ruffini, R., Vereshchagin, G. V., Xue, S.-S., {\it Physics Reports}, 2010, 487, 1-4, 1-140

\bibitem{belvedere12} Belvedere, R., Pugliese, D., Rueda, J. A., Ruffini, R., Xue, S.-S., {\it Nuclear Physics A} 2012, 883, 1-24

\bibitem{chatzisavvas09} Chatzisavvas, K. Ch.; Psonis, V. P., Panos, C. P. and Moustakidis, Ch. C., {\it Physics Letters A} 2009, 373, 3901-3909



\end{thebibliography}
\end{document}